\documentclass[a4paper, 11pt]{article}

\usepackage{graphicx}
\usepackage{amsmath}
\usepackage{amsfonts}
\usepackage{amssymb}
\usepackage{authblk}

\usepackage{amsthm,amsmath,natbib}
\RequirePackage{hyperref}
\usepackage{multirow}
\usepackage{amsmath,amssymb}

\usepackage{pdfpages}
\usepackage{caption}
\numberwithin{equation}{section}
\theoremstyle{plain}

\emergencystretch=\maxdimen
\usepackage{tikz}
\usetikzlibrary{arrows,backgrounds,snakes,patterns}
\usetikzlibrary{shapes,arrows,chains}
\usepackage{verbatim}
\usepackage{booktabs}

\usepackage[flushleft]{threeparttable}

\begin{document}
\pdfoutput=1
\thispagestyle{empty} 

\title{Bayesian significance test for discriminating between survival distributions}

\author[1]{Cachimo Combo Assane \footnote{cachimo.assane@gmail.com}}
\author[1]{Basilio de Bragan\c{c}a Pereira \footnote{basilio@hucff.ufrj.br}}
\author[2]{Carlos Alberto de Bragan\c{c}a Pereira \footnote{cpereira@ime.usp.br}}

\affil[1]{Universidade Federal do Rio de Janeiro (UFRJ), Rio de Janeiro, Brazil}
\affil[2]{Universidade de S\~{a}o Paulo (USP), S\~{a}o Paulo, Brazil}

\maketitle

\begin{abstract}
An evaluation of FBST, Fully Bayesian Significance Test, restricted to survival models is the main objective of the present paper. A Survival distribution should be chosen among the tree celebrated ones, lognormal, gamma, and Weibull.  For this discrimination, a linear mixture of the three distributions, for which the mixture weights are defined by a Dirichlet distribution of order three, is an important tool: the FBST is used to test the hypotheses defined on the mixture weights space.  Another feature of the paper is that all three distributions are reparametrized in that all the six parameters -- two for each distribution -- are written as functions of the mean and the variance of the population been studied. Note that the three distributions share the same two parameters in the mixture model. The mixture density has then four parameters, the same two for the three discriminating densities and two for the mixture weights. Some numerical results from simulations with some right-censored data are considered.  The lognormal-gamma-Weibull model is also applied to a real study with dataset being composed by patient's survival times of patients in the end-stage of chronic kidney failure subjected to hemodialysis procedures; data from Rio de Janeiro hospitals.  The posterior density of the weights indicates an order of the mixture weights and the FBST is used for discriminating between the three survival distributions \\ \\

\noindent \emph{Keywords}: Model choice; Separate Models; Survival distributions; Mixture model; Significance test; FBST
\end{abstract}
\section{Introduction}\label{sec1}

In many scientific disciplines, researchers are constantly faced with the fundamental problem of choosing among alternative statistical models. The Neyman-Pearson theory of hypothesis testing applies only if the models belong to the same family of distributions. Alternatively, special procedures are required if the models belong to families that are separate (or non-nested) in the sense that an arbitrary member of one family cannot be obtained as a limit of members of the other. The set of separate families of probability distributions includes the ones used here: lognormal, gamma, and Weibull models \citep{Pereira81,Araujo07,Pereira17} which have been used widely to describe survival data \citep{Lawless02,LeeWang}.

A considerable amount of research on separate families of hypotheses has been realized since the fundamental work of \citet{Cox61, Cox62}, who first dealt with the problem. For reviews and references, see \citet{Araujo05}; \citet{Araujo07};  and \citet{Pereira17}.

The Fully Bayesian Significance Test (FBST) introduced by \citet{PereiraS} is an alternative test to the ones that are based on Bayes factor or on the classical p-value; mostly for the case of precise hypotheses. The basis for the FBST is an index known as e-value (e stands for evidence) that measures the inconsistency of the hypothesis. For this, it considers the tangent set, $T$; the set of all parameter values for which their posterior density values are greater than the values of the posterior densities of all points that attend the hypothesis. For reviews and further references on FBST, see \cite{PereiraSW} and \citet{SternP14}. For a few interesting applications illustrating the use of e-values and the FBST to practical problems, see \citet{Diniz12}, \citet{Lauretto03}, \citet{Lauretto07}, and \citet{PereiraS}.

In the present work, we consider the FBST for discriminating between the lognormal, gamma and Weibull distributions. We formulate this problem in the context of linear mixture model, as suggested by \citet{Cox61}. It means that, the models under comparison are considered as components of a finite mixture model. The FBST is used for testing hypotheses defined on the mixture weights space.  The e-value is the complementary of the posterior probability of the tangent set $T$; $ev = 1-Pr(T|Data)$,

Additionally, the density functions of the mixture components are reparametrized in terms of the mean  $\mu$ and the variance $\sigma^2$ of the population. Hence, the models under discrimination share common parameters \citep{kamary14,Pereira17}. A standard Bayesian approach to finite mixture models is to consider different pairs of parameters for each of these models and to adopt independent prior distributions for each pair of parameters and a Dirichlet prior on the mixture weights \citep{Lauretto05,Lauretto07}. However, since the comparison between the models is based on the same dataset and on the same sample, we believe that it would be inappropriate to consider different means and variances for these models. Note that this reparameterization reduces the number of the parameters to be estimated: in our case, including the weights, from eight to only four.

To illustrate the procedure, numerical results based on simulated right-censored survival times were considered. Also, a real example is introduced to use the lognormal-gamma-Weibull mixture model to the dataset of patients, from Rio de Janeiro hospitals, with end-stage chronic kidney failure who received hemodialysis.

Section \ref{sec2} presents a brief review of basic concepts and notation for survival analysis. The parametric distributions used in this paper are also described. Section \ref{sec3} reviews the basic concepts o FBST. Section \ref{sec4} discusses the FBST formulation for discriminating between survival distributions in the context of mixture models. Section \ref{sec5} presents the results of the simulation study. Section \ref{sec6} is about the use of the lognormal-gamma-Weibull on the real dataset. Final remarks are presented in Section \ref{sec7}.

\section{Survival analysis} \label{sec2}

\subsection{Basic concepts and notation}
Survival analysis is concerned with the analysis of time to occurrence of a certain event of interest, such as failure, death, relapse or development of a given disease.

Let $T$ be a non-negative random variable representing the time until some event of interest. There are three functions of primary interest used to characterize the distribution of $T$, namely the survival function, the probability density function and the hazard function \citep{LeeWang}.

The survival function, denoted by $S(t)$, is defined as the probability that an individual survives beyond time $t$:
\begin{equation}\label{surv}
S(t)=P(T>t)=1-F(t), \ \ \mbox{for} \ \ t>0,
\end{equation}
where $F(t)$ is the distribution function of $T$. Note that $S(t)$ is a nonincreasing continuous function of time $t$ with $S(0)=1$ and $S(\infty)=\displaystyle\lim_\infty S(t)=0$.

The probability  density function, denoted by $f(t)$, is the probability of failure in a small interval per unit time. It can be expressed as
\begin{equation}\label{densfun}
f(t)=\frac{dF(t)}{dt}=\frac{d\{1-S(t)\}}{dt}=-\frac{dS(t)}{dt}.
\end{equation}
The hazard function, denoted by $h(t)$, represents the probability of failure during a very small time interval, assuming that the individual has survived to the beginning of the interval:
\begin{equation}\label{hazfun}
h(t)=\displaystyle\lim_{\Delta t \to 0}\frac{\mbox{P}(t\leq T< t+ \Delta t|T \geq t)}{\Delta t}=\frac{f(t)}{S(t)}.
\end{equation}
This function is also known as the conditional failure rate. The cumulative hazard function is defined as

\begin{equation}\label{cumhazfun}
H(t)=\displaystyle \int_{0}^{t}h(u)d(u).
\end{equation}
Therefore, when $t=0$ then, $S(t)=1$ and $H(t)=0$; and when $t=\infty$ then, $S(t)=0$ and $H(t)=\infty$.

\subsection{Parametric survival distributions}
In this paper, we consider the the FBST for discriminating between the lognormal, gamma and Weibull distributions which are most frequently used in modeling survival data \citep{Lawless02,LeeWang}. The probability density functions, the survival functions and the hazard functions of these distributions are highlighted below.
\begin{itemize}
\item [i)] Let $T$ be a lognormal random variable with parameters $\alpha=(\alpha_1,\alpha_2)$, denoted by $T\sim LN(\alpha_1,\alpha_2)$,
 \begin{align*}
  f_L(t|\alpha)&=\frac{1}{t\sqrt{2\pi\alpha_2}}\exp\left\{-\frac{(\log t-\alpha_1)^2}{2\alpha_2}\right\},  \ \ -\infty<\alpha_1<\infty, \alpha_2, t>0;&\\
  S_{L}(t|\alpha)&=\frac{1}{\sqrt{2\pi\alpha_2}}\int_{t}^{\infty}\frac{1}{t}\exp\left\{-\frac{(\log t-\alpha_1)^2}{2\alpha_2}\right\}dy&\\
  &=1-\Phi\left[\frac{(\log t-\alpha_1)}{\sqrt{\alpha_2}}\right];&\\
  h_{L}(t|\alpha)&= \frac{f_{LN}()}{S_{LN}()}.&
  \end{align*}
\item [ii)] If $T$ has a Gamma distribution with parameters $\gamma=(\gamma_1,\gamma_2)$, denoted by $T\sim G(\gamma_1,\gamma_2)$, then
  \begin{align*}
   f_G(t| \gamma)&=\frac{1}{\Gamma(\gamma_2)\gamma_1^{\gamma_2}}t^{\gamma_2-1}\exp\left\{-\frac{t}{\gamma_1}\right\}, \  \gamma_1, \gamma_2,  t>0;&\\ \vspace*{0.8cm}
   S_G(t|\gamma)&=1-\displaystyle\int_{0}^{t}\frac{1}{\Gamma(\gamma_2)\gamma_1^{\gamma_2}}u^{\gamma_2-1}\exp\left\{-\frac{u}{\gamma_1}\right\} du;&\\ \vspace*{0.8cm}
   h_G(t|\gamma)&= \frac{f_G()}{S_G()}.&
  \end{align*}
\item [iii)] If $T$ has a Weibull distribution with parameters $\beta=(\beta_1,\beta_2)$, denoted by $T\sim W(\beta_1,\beta_2)$, then
  \begin{align*}
   f_W(t|\beta)&=\frac{\beta_2}{\beta_{1}^{\beta_2}}t^{\beta_2-1}\exp\left\{-\left(\frac{t}{\beta_1}\right)^{\beta_2}\right\}, \  \beta_1, \beta_2,  t>0;&\\\vspace*{0.8cm}
   S_W(t|\beta)&=\exp\left\{-\left(\frac{t}{\beta_1}\right)^{\beta_2}\right\};&\\ \vspace*{0.8cm}
   h_W(t|\beta)&=\frac{\beta_2}{\beta_{1}^{\beta_2}}t^{\beta_2-1}.&
\end{align*}
\end{itemize}

\section{Fully Bayesian Significance Test (FBST)}\label{sec3}

The FBST of \citet{PereiraS}, which is reviewed in \citet{PereiraSW}, is a Bayesian version of significance testing, as considered by \citet{Cox77} and \citet{Kempt76}, for precise (or sharp) hypotheses.

First, let us consider a real parameter $\theta$, a point in the parameter space $\Theta\subset\Re$, and an observation $y$ of the random variable $Y$. A frequentist looks for the set $I\in\Re$ of sample points that are at least as inconsistent with the hypothesis as $y$ is. A Bayesian looks for the tangential set $T(y)\subset\Theta$ \citep{PereiraSW}, which is a set of parameter points that are more consistent with the observed $y$ than the hypothesis is. An example of a sharp hypothesis in a parameter space of the real line is of the type $H: \theta=\theta_0$. The evidence value in favor of $H$ for a frequentist is the usual p-value, $P(Y\in I|\theta_0)$, whereas for a Bayesian, the evidence in favor of $H$ is the e-value, $ev=1-\mbox{Pr}(\theta\in T(y)|y)$.

In the general case of multiple parameters, $\Theta\subset\Re^k$, let the posterior distribution for $\theta$ given $y$ be denoted by $q(\theta|y)\propto \pi(\theta)L(y,\theta)$, where $\pi(\theta)$ is the prior probability density of $\theta$ and $L(y,\theta)$ is the likelihood function. In this case, a sharp hypothesis is of the type $H:\theta\in\Theta_H\subset\Theta$, where $\Theta_H$ is a subspace of smaller dimension than $\Theta$. Letting $\displaystyle\sup_H$ denote the supremum of $\Theta_H$, we define the general Bayesian evidence and the tangential set, $T(y)$, as follows:
\begin{equation}\label{tangent}
q^{*}=\displaystyle\sup_Hq(\theta|y) \ \ \mbox{and} \ \ T(y)= \{\theta: q(\theta|y)>q^{*} \}.
\end{equation}
The Bayesian  evidence value against $H$ is the posterior probability of $T(y)$,
\begin{equation}\label{ev}
\overline{ev}=\mbox{Pr}(\theta\in T(y)|y)=\int_{T(y)}q(\theta|y)d\theta; \ \ \mbox{consequently}, \ \ ev=1-\overline{ev}.
\end{equation}

It is important to note that evidence that favors $H$ is not evidence against the alternative, $\overline{H}=\Theta\setminus H$, because it is not a sharp hypothesis. This interpretation also holds for p-values in the frequentist paradigm. As in \citet{PereiraSW}, we would like to point out that this Bayesian significance index uses only the posterior distribution, with no need for additional artifacts such as the inclusion of positive prior probabilities for the hypotheses or the elimination of nuisance parameters. The computation of the e-values does not require asymptotic methods, and the only technical tools needed are numerical optimization and integration methods.

\section{Mixture of survival models}\label{sec4}

Let us consider a dataset $y=\{y_1, \ldots, y_n\}$ and $m$ alternative parametric survival distributions with densities $f_1(y|\psi_1),f_2(y|\psi_2),\ldots, f_m(y|\psi_m)$. Here, $\psi_k, k=1,\ldots,m$, are unknown (vector) parameters and the families of distributions are separate. The problem of interest is to measure the evidence in favor of each model for fitting the dataset. As suggested by \citet{Cox61}, we can consider a general model including all candidate distributions where the choice of a specific distribution is a special case. In this work, we formulate the FBST for the linear mixture of the survival models as a selection procedure. Denoting $\boldsymbol{\theta}=(\psi_1,\ldots,\psi_m, \boldsymbol{p})$, the density function for $m-$component mixture model is
\begin{equation} \label{mixture model}
f(y_j|\boldsymbol{\theta})=p_1f_1(y_j|\psi_1)+\ldots+p_mf_m(y_j|\psi_m) \ \ p_k\geq0, \ \displaystyle\sum_{k=1}^{m}p_k=1.
\end{equation}
where $\boldsymbol{p}=(p_1,\ldots,p_m)$ is the vector of the mixture weights.

In the presente work, the density functions of the mixture components in (\ref{mixture model}) are reparametrized in terms of the mean $\mu$ and the variance $\sigma^2$ of the population. Hence, the models under comparison share common parameters \citep{kamary14,Pereira17}. The main reason for this reparametrization is that, since the comparison between the models is based on the same dataset and on the same sample, we believe that it would be inappropriate to consider different means and variances for these models as is commonly performed in traditional Bayesian approach to finite mixture model. Therefore, we have $\boldsymbol{\theta}=(\mu,\sigma^2, \boldsymbol{p})$ denoting all parameters of the mixture model, where $\mu$ and $\sigma^2$ are the connecting parameters, with $\boldsymbol{p}$ corresponding to the vector of the mixture weights.

Assuming that the $y_i$ are conditionally (on the parameter) independent, the likelihood function is defined as
\begin{equation} \label{verossP}
L(y,\boldsymbol{\theta})=\displaystyle\prod_{j=1}^{n}\sum_{k=1}^{m}p_kf_k(y_j|\mu,\sigma).
\end{equation}

The families of distributions considered include the lognormal, gamma and Weibull models. Hence, the relationship between the parameters of these models through the $\mu$ and $\sigma^2$ is described as follows.

\begin{itemize}
  \item [(i)] Let $y$ be a $\mbox{lognormal}(\alpha_1,\alpha_2), \alpha_1\in \ \mathbb{R} \ \mbox{and} \alpha_2>0$, with probability density function
  \begin{equation*}
  f_L(y|\alpha_1,\alpha_2)=\frac{1}{y\sqrt{2\pi\alpha_2}}\exp\left\{-\frac{(\log y-\alpha_1)^2}{2\alpha_2}\right\}.
  \end{equation*}
 We then have
\begin{equation}\label{rep_lnormal}
\left\{
  \begin{array}{ll}
 \vspace*{0.3 cm}
    \mu= E(y|\alpha_1,\alpha_2)=\mbox{e}^{\alpha_1+\alpha_2/2}\\
    \sigma^2=Var(y|\alpha_1,\alpha_2)=(\mbox{e}^{\alpha_2}-1)\mbox{e}^{2\alpha_1+\alpha_2}
  \end{array}
\right. \Rightarrow\left\{
                     \begin{array}{ll}
                         \vspace*{0.3 cm}
                       \alpha_1=\log\frac{\mu^2}{\sqrt{\mu^2+\sigma^2}}\\
                       \alpha_2=\sqrt{\log\frac{\mu^2+\sigma^2}{\mu^2}}.
                     \end{array}
                   \right.
\end{equation}

  \item [(ii)] Let $y$ be a $\mbox{gamma}(\gamma_1,\gamma_2), \gamma_1>0 \ \mbox{and} \ \gamma_2>0$, with probability density function
  \begin{equation*}
  f_G(y| \gamma_1,\gamma_2)=\frac{1}{\Gamma(\gamma_2)\gamma_1^{\gamma_2}}y^{\gamma_2-1}\exp\left\{-\frac{y}{\gamma_1}\right\}.
  \end{equation*}
Therefore
\begin{equation}\label{rep_gama}
\left\{
  \begin{array}{ll}
 \vspace*{0.3 cm}
    \mu= E(y|\gamma_1,\gamma_2)=\gamma_1\gamma_2\\
    \sigma^2=Var(y|\gamma_1,\gamma_2)=\gamma_2\gamma_{1}^{2}
  \end{array}
\right. \Rightarrow\left\{
                     \begin{array}{ll}
                         \vspace*{0.3 cm}
                       \gamma_1=\frac{\sigma^2}{\mu}\\
                       \gamma_2=\frac{\mu^2}{\sigma^2}.
                     \end{array}
                   \right.
\end{equation}

  \item [(iii)] When $y\sim\mbox{Weibull}(\beta_1,\beta_2),\beta_1>0 \ \mbox{and} \ \beta_2>0$, with probability density function
\begin{equation*}
 f_W(y|\beta_1,\beta_2)=\frac{\beta_2}{\beta_{1}^{\beta_2}}y^{\beta_2-1}\exp\left\{-\left(\frac{y}{\beta_1}\right)^{\beta_2}\right\},
 \end{equation*}
then
\begin{align}\label{rep_weibull}
&\left\{
  \begin{array}{ll}
 \vspace*{0.3 cm}
    \mu= E(y|\beta_1,\beta_2)=\beta_1\Gamma(1+1/\beta_2)\\
    \sigma^2=Var(y|\beta_1,\beta_2)=\beta_{1}^{2}\Gamma(1+2/\beta_2)-\beta_{1}^{2}\Gamma^{2}(1+1/\beta_2)
  \end{array}
\right. \nonumber\\ & \nonumber\\ \Rightarrow&\left\{
                     \begin{array}{ll}
                         \vspace*{0.3 cm}
                       \beta_1=\frac{\mu}{\Gamma(1+1/\beta_2)}\\
                       2\log\Gamma(1+1/\beta_2)-\log\Gamma(1+2/\beta_2)+\log\frac{\mu^2+\sigma^2}{\mu^2}=0.
                     \end{array}
                   \right. &
\end{align}
\end{itemize}
In order to find $\beta_2$, the Newton-Rapson method can be used to solve the nonlinear equation. Here, we use the \verb"nleqslv" function in the \verb"R" package of the same name.

A special feature of survival data is that survival times are frequently censored. The survival time of an individual is said to be censored when the event of interest has not been observed for that individual, but is known only to occur in a certain period of time. There are various categories of censoring, such as right censoring, left censoring and interval censoring (see \citet{Klein} for more details). In this paper, we restrict ourselves to data in which the survival times are subject to right censoring, which is the most common censoring mechanism in medical research.

In the model for right-censored data, it is convenient to consider the following notation. Each individual $j$ is assumed to have an event time $T_j$ and a censoring time $C_j$. The observations consist of $(y_1,\delta_1), (y_2,\delta_2), \ldots,(y_n,\delta_n)$, where $y_j=\min\{T_j, C_j\}$ and $\delta_j=\mathbb{I}(T_j\leq C_j)$, indicating whether $T_j$ was observed ($\delta_j=1$) or not ($\delta_j=0$).

Note that the likelihood function given by (\ref{verossP}) is for uncensored (or exact) observations. Assuming noninformative censoring, i.e, independence between $T_j$ and $C_j$, then, the likelihood function for right-censored observations is

\begin{align}
L(y,\boldsymbol{\theta})&=\displaystyle\prod_{j}^{n}f(y_j,\delta_j|\boldsymbol{\theta})& \nonumber\\
           &\propto\displaystyle\prod_{j}^{n}[f(y_j|\boldsymbol{\theta})]^{\delta_j}[S(y_j|\boldsymbol{\theta})]^{1-\delta_j}& \nonumber\\
           &\propto \displaystyle\prod_{j}^{n}\left[\displaystyle\sum_{k=1}^{m} p_kf_k(y_j|\mu,\sigma)\right]^{\delta_j} \left[\sum_{k=1}^{m} p_kS_k(y_j|\mu,\sigma)\right]^{1-\delta_j},&
\end{align}
where, $S_k$ is the survival function associated with the mixture component $k$.

Assuming independence, the joint prior density function of $\boldsymbol{\theta}=(\mu,\sigma^2,\boldsymbol{p})$ is given by $\pi(\boldsymbol{\theta})=\pi_1(\boldsymbol{p})\pi_2(\mu)\pi_3(\sigma^2)$. Therefore, according to the Bayesian paradigm, the posterior density of $\boldsymbol{\theta}$ is
\begin{equation}\label{joint posterior}
f(\boldsymbol{\theta}|y)\propto L(y,\boldsymbol{\theta})\pi(\boldsymbol{\theta}).
\end{equation}

In this paper, the prior distributions for the connecting parameters, $\mu$ and $\sigma^2$, are assumed to be independent gamma distributions, both with a mean of one and a variance of 100, that is, $\mu, \sigma^2\sim gamma(0.01,100)$ \citep{Pereira17}. For the mixture weights, we use a Dirichlet prior, $\boldsymbol{p}\sim Dir(1,1,1)$ when all families of models are considered ($m=3$) or a Beta prior with parameters (1,1) (uniform$(0,1)$) for any combination of $m=2$.

In order to measure the evidence in favour of each model, the hypotheses on the mixture weights are tested \citep{kamary14,Pereira17}.

The hypothesis specifying that $y$ has the density function $f_k(y|\psi_k)$ is equivalent to
\begin{equation}\label{hp1}
H_k: p_k=1 \wedge p_i=0, i\neq k.
\end{equation}

On the other hand, the hypothesis that $y$ has not the density $f_k(y|\psi_k)$ is equivalent to
\begin{equation}\label{hp0}
H: p_k=0 \wedge \displaystyle\sum_{i\neq k} p_i=1.
\end{equation}

The alternative hypotheses to (\ref{hp1}) and (\ref{hp0}) are $A_k: p_k<1$ and $A_k: p_k>0$, respectively, which are not sharp anyway.

The FBST procedure is used to test $H_k,k=1,\ldots,m$, according to the expressions (\ref{tangent}) and (\ref{ev}). For the optimization step, we used the conjugate gradient method \citep{Fletcher}. In order to perform the integration over the posterior measure, we used an Adaptive Metropolis Markov chain Monte Carlo algorithm (MCMC) of \citet{Haario01}.

In this paper, the implementation of the Bayesian models is carried out using \verb"LaplacesDemon" \verb"R" package. The \verb"LaplacesDemon" is an open-source package that provides a complete environment for simulation in Bayesian inference \citep{LCC16}.

\section{Simulations}\label{sec5}
In this section we present some numerical results based on simulated right-censored survival times in order to evaluate the performance of the FBST for discriminating between the survival distributions via lognormal-gamma-Weibull mixture model (LGW). The main purpose is to measure the convergence rate of correct decisions, concerning the identification of the true model used to generate the survival times $T$.

The simulations of this paper were performed on a Intel(R) Core(TM) i7-5500U CPU@ 2.40GHz computer.

\subsection{Simulation scheme of sample points}

Let $H_L$, $H_G$ and $H_W$ be the hypotheses specifying the probability density functions of the lognormal, gamma and Weibull distributions, respectively. From each distribution, we generate $200$ samples of sizes $n=100$, $200$, $300$, and $500$. Each sample contain a desired proportion of right-censored observations.

The steps used to simulate a sample, $y$, of size $n$, in which part of the observations is right-censored, are shown below. For this example, we assume that the true survival times has a lognormal distribution.
\begin{enumerate}
  \item Assign values to parameters $\mu$ e $\sigma^2$;
  \item Calculate the lognormal parameters $(\alpha_1, \alpha_2)$ using the expressions (\ref{rep_lnormal});
  \item For $j=1,\ldots, n$,
  \begin{itemize}
    \item Generate the survival time $T_j$ from  $lognormal(\alpha_1, \alpha_2)$;
    \item Generate the right-censoring time $C_j$ from a exponential distribution, i.e, $C_j\sim Exp(\lambda)$, where the parameter $\lambda$ is chosen such that approximately a desired percentage of simulated observations are right-censored;
    \item Obtain the observed time $y_j=\min\{T_j, C_j\}$
    \item Create an indicator random variable $\delta_j=\mathbb{I}(T_j\leq C_j)$
  \end{itemize}
\end{enumerate}
Using this generated sample, we obtain the posterior samples for the mixture parameters from \emph{Adaptive Metropolis} algorithm and we use the FBST to calculate the evidence measures in favor of each model.

The value for the censoring distribution parameter, $\lambda$, is determined by numerical methods \citep{Wan17}. We let $p_c$ denote the right-censoring probability. We suppose that the censoring time $C$ has exponencial density function $g(c|\lambda)$ and the independence assumption between $T$ and $C$ holds. In order to simulate a sample with approximately $p_c\%$ of right-censored observations, the value of $\lambda$ is obtained by solving the following equation:
\begin{align}\label{eq.cens}
p_c&=\mbox{Pr}(\delta=0|\lambda, \mu, \sigma^2)& \nonumber\\
   &=\mbox{Pr}(C\leq T\leq \infty, 0\leq C\leq\infty)&\nonumber\\
   &=1-\mbox{Pr}(0\leq T\leq C, 0\leq C\leq\infty)&\nonumber\\
   &=1-\displaystyle\int_{0}^{\infty}g(c|\lambda)\displaystyle\int_{0}^{c}f_L(t|\mu,\sigma)dtdc&\nonumber\\
   &=1-\displaystyle\int_{0}^{\infty}g(c|\lambda)F_L(c|\mu,\sigma)dc,&
\end{align}
where $f_L$ and $F_L$ are the lognormal probability density and distribution functions of survival times, respectively.

For generating right-censored survival times from the gamma and Weibull distributions, an analogous procedure to that used for the lognormal distribution is employed.

\subsection{Criteria for evaluating the performance of the FBST}

In order to evaluate the performance of the FBST on selecting the true distribution used to generate the survival times, we have compared the measures of evidence in favor of the hypotheses $H:p_k=0$ and $H:p_k=1$, $k=L,G,W$, where $p_k$ are respectively the mixture weights associated with the lognormal, gamma and Weibull components in the LGW mixture model.

For instance, suppose again that the true survival time has a lognormal distribution. We consider that the FBST has made a correct choice on the LGW model, if the evidence in favor of $H:p_L=0$ is less than that in favor of $H:p_G=0$ and $H:p_W =0$, and the evidence in favor of $ H:p_L= 1$ is greater than that in favor of $H:p_G=1 $ e $H:p_W=1$.

The calculation of the proportions of correct decisions made by FBST is based on $200$ replicates. In these simulations, we have assigned $\mu=20$ and $\sigma^2=50$. The FBST procedure is evaluated considering the samples with different censoring percentages: $10\%$, $30\%$ and $50\%$.

\subsection{Simulation results}
Table \ref{LGW simulation} presents the mean of the estimates for the LGW mixture model parameters and the percentages of correct decisions made by FBST on selecting the true distribution used to generate the survival times. It is observed that, regardless of the distribution used for generating the survival times and the sample sizes, the estimates for the mean $\mu$ are very close to each other and to the true value of the parameter. For the estimates of the variance $\sigma^2$, we observe a variation between them but, in general, they approach the true value of the parameter as the sample size increases.

It is observed that the FBST presents a high performance on identifying the Weibull distribution as the true data generation process and low performance on identifying the gamma distribution. This happens because, regarding the parameters chosen for these simulations, the gamma and lognormal densities are very similar. The general pattern of the simulation results shows that the FBST achieves good performance even for samples with $50\%$ right-censoring.

\begin{table}[htp!]
\caption{\small Mean of estimates for LGW model parameters and percentages of correct decisions made by FBST on selecting the true distribution used to generate the survival times, using samples with different right-censoring percentages} \label{LGW simulation}
{\small
\begin{tabular}{ccccccccc} \hline
\multirow{2}{*}{$\%$ of Rc$^{\dag}$} &\multirow{2}{*}{Model}&\multirow{2}{*}{$n$}&$\mu$&$\sigma^2$&$p_L$&$p_G$&$p_W$&\multirow{2}{*}{$\%$ of Cd$^{\ddag}$} \\ \cline{4-8}
                         &                    &                       &$20$&$50$&-&-&-&\\ \hline
\multirow{12}{*}{$10$}&\multirow{4}{*}{Lognormal}&$100$&$19.92$&$48.82$&$0.48$&$0.36$&$0.16$&$84$ \\
                      &                          &$200$&$20.02$&$48.87$&$0.59$&$0.30$&$0.11$&$84$ \\
                      &                          &$300$&$19.96$&$48.43$&$0.63$&$0.28$&$0.09$&$84$ \\
                      &                          &$500$&$19.97$&$48.14$&$0.69$&$0.25$&$0.07$&$93$ \\ \cline{3-9}
                      &\multirow{4}{*}{Gamma}     &$100$&$20.07$&$52.95$&$0.38$&$0.36$&$0.26$&$45$ \\
                      &                          &$200$&$20.01$&$50.60$&$0.38$&$0.41$&$0.21$&$53$\\
                      &                          &$300$&$20.06$&$50.90$&$0.36$&$0.44$&$0.20$&$57$\\
                      &                          &$500$&$20.05$&$50.91$&$0.34$&$0.48$&$0.18$&$69$\\ \cline{3-9}
                      &\multirow{4}{*}{Weibull}  &$100$&$20.17$&$52.06$&$0.19$&$0.26$&$0.55$&$86$ \\
                      &                          &$200$&$20.04$&$51.27$&$0.15$&$0.22$&$0.63$&$94$\\
                      &                          &$300$&$19.76$&$50.24$&$0.12$&$0.18$&$0.70$&$100$\\
                      &                          &$500$&$19.97$&$51.26$&$0.09$&$0.13$&$0.78$&$100$ \\ \hline
 \multirow{12}{*}{$30$}&\multirow{4}{*}{Lognormal}&$100$&$20.03$&$49.20$&$0.47$&$0.35$&$0.19$&$48$ \\
                      &                          &$200$&$20.01$&$48.71$&$0.55$&$0.33$&$0.13$&$63$ \\
                      &                          &$300$&$20.02$&$48.20$&$0.59$&$0.31$&$0.10$&$70$ \\
                      &                          &$500$&$19.97$&$47.06$&$0.64$&$0.28$&$0.08$&$86$ \\ \cline{3-9}
                      &\multirow{4}{*}{Gamma}     &$100$&$20.13$&$53.03$&$0.38$&$0.36$&$0.26$&$35$ \\
                      &                          &$200$&$19.96$&$50.67$&$0.40$&$0.38$&$0.22$&$47$\\
                      &                          &$300$&$20.12$&$55.09$&$0.41$&$0.41$&$0.18$&$51$\\
                      &                          &$500$&$20.00$&$50.77$&$0.35$&$0.47$&$0.18$&$70$\\ \cline{3-9}
                      &\multirow{4}{*}{Weibull}  &$100$&$20.07$&$54.25$&$0.21$&$0.28$&$0.51$&$81$ \\
                      &                          &$200$&$20.11$&$52.35$&$0.16$&$0.23$&$0.61$&$92$\\
                      &                          &$300$&$19.97$&$50.77$&$0.15$&$0.21$&$0.64$&$100$\\
                      &                          &$500$&$19.97$&$50.78$&$0.10$&$0.15$&$0.75$&$100$ \\ \hline
 \multirow{12}{*}{$50$}&\multirow{4}{*}{Lognormal}&$100$&$19.98$&$48.14$&$0.45$&$0.35$&$0.20$&$40$ \\
                      &                          &$200$&$19.91$&$45.79$&$0.50$&$0.35$&$0.15$&$50$ \\
                      &                          &$300$&$19.97$&$47.14$&$0.56$&$0.33$&$0.11$&$65$ \\
                      &                          &$500$&$19.92$&$46.82$&$0.63$&$0.29$&$0.08$&$78$ \\ \cline{3-9}
                      &\multirow{4}{*}{Gamma}     &$100$&$20.06$&$53.77$&$0.36$&$0.36$&$0.28$&$31$ \\
                      &                          &$200$&$19.99$&$51.17$&$0.37$&$0.38$&$0.25$&$43$\\
                      &                          &$300$&$20.10$&$52.55$&$0.37$&$0.41$&$0.22$&$47$\\
                      &                          &$500$&$20.08$&$51.65$&$0.40$&$0.43$&$0.17$&$57$\\ \cline{3-9}
                      &\multirow{4}{*}{Weibull}  &$100$&$20.27$&$58.79$&$0.24$&$0.30$&$0.46$&$74$ \\
                      &                          &$200$&$20.09$&$54.65$&$0.18$&$0.26$&$0.56$&$90$\\
                      &                          &$300$&$20.05$&$53.08$&$0.15$&$0.23$&$0.62$&$95$\\
                      &                          &$500$&$19.92$&$51.21$&$0.13$&$0.19$&$0.68$&$95$ \\ \hline
                      \multicolumn{9}{l}{\small$\dag$ percentage of right-censoring} \\
                       \multicolumn{9}{l}{\small$\ddag$ percentage of correct decision}
\end{tabular}
}
\end{table}

\section{Application: Choice of a survival model for patients with end-stage kidney disease}\label{sec6}

\subsection{Dataset}

The dataset used in this paper refers to a cohort study of 473 patients with end-stage chronic kidney failure who received hemodialysis (HD) in four centers in the State of Rio de Janeiro, Brazil. The patients were followed up $11$ years. The observed time for each patient was the number of months from admission to hemodialysis until death or the end of the observation period (kidney transplant or end of the study) which indicates a right-censored survival time. For a complete description of this dataset, see \citet{Mauro}.

In this paper, our main interest is to apply the LGW model to the survival data for HD patients and use the FBST procedure to examine the mixture parameters in order to choose the parametric distribution that best fits the observed data. But before that, we have performed pairwise comparisons by fitting the lognormal-Weibull, lognormal-gamma, and gamma-Weibull mixture models.

\subsection{Results}

\begin{table}[htp!]
\caption{\small Measures of evidence provided by HD data} \label{HD}
\begin{tabular}{lccc} \hline
\multirow{3}{*}{Comparison}&\multirow{3}{*}{Null hypothesis}&\multicolumn{2}{c}{Evidence in favor of null hypothesis} \\ \cline{3-4}
                           &  &\multirow{2}{*}{e-value}& \multirow{2}{*}{p-value$^{*}$} \\
                           &  &                        &          \\ \hline
\multirow{2}{*}{$H_L \times H_W$}&$H_L$ &$0.874$ &$0.404$ \\
                              &$H_W$ & $0.043$& $0.004$ \\ \hline
\multirow{2}{*}{$H_L \times H_G$}&$H_L$&$0.901$ & $0.446$\\
                                 &$H_G$& $0.757$& $0.277$\\ \hline
 \multirow{2}{*}{$H_G \times H_W$}&$H_G$& $1.000$& $1.000$ \\
                                  &$H_W$& $0.113$& $0.015$   \\ \hline
 \multicolumn{3}{l}{\small*p-value calculated according to \citet{Diniz12}}
\end{tabular}
\end{table}
The measures of evidence provided by HD data in favor of the three models concerning the pairwise comparisons are presented in Table \ref{HD}. For the comparison between the lognormal and Weibull distributions, the FBST indicates to choose the lognormal model since the e-values $ev(H_L)=0.874$ and $ev(H_W)=0.043$. For selecting between the lognormal and gamma distributions, the evidence measures indicate that both models provide good fit to the dataset. Nevertheless, also we would prefer to choose the lognormal model which is the most plausible. The results of the tests for comparison between the gamma and Weibull distributions indicate that the Weibull distribution does not provide reasonable fit to the dataset.

\vspace*{.3 cm}
\noindent \textbf{Discrimination based on the LGW mixture model}
\vspace*{.3 cm}

In order to test simultaneously the three hypotheses, we have applied the the LGW model,
\begin{equation} \label{LGW}
f(y|\boldsymbol{p},\mu,\sigma)=p_1f_L(y|\mu,\sigma)+p_2f_{G}(y| \mu, \sigma)+p_3f_W(y|\mu,\sigma),
\end{equation}
to the HD data.

The estimates for the parameters of the model (\ref{LGW}) are presented in Table \ref{EstLGW}. Here, SD, $2.5\%$ and $97.5\%$ denote the standard deviation, the $2.5$th and the $97.5$th percentiles of the posterior distribution of the LGW parameters, respectively. Both the classical and the Bayesian measures of evidence, presented in Table \ref{testpLGW}, indicate that neither the gamma and Weibull models should be considered because the null hypotheses $H: p_2=0$ e $H: p_3=0$ are not rejected. Consequently, among the three models, the lognormal model is the most appropriate for modeling HD data.

\begin{table}[htp!]
\caption{\small Summary of the posterior distribution of the LGW parameters}\label{EstLGW}
\begin{tabular}{lcccccc} \hline
Parameter&Mean&SD&$2.5\%$&Median&$97.5\%$ \\\hline
$p_1\mbox{-lognormal}$&$0.574$ &$0.299$&$0.028$&$0.656$&$0.957$ \\
$p_2\mbox{-gamma}$&$0.283$ &$0.256$&$0.005$&$0.187$&$0.845$ \\
$p_3\mbox{-Weibull}$&$0.143$ &$0.157$&$0.003$&$0.087$&$0.606$ \\
$\mu$&$18.537$&$1.441$&$15.819$&$18.561$&$21.079$\\
$\sigma^2$&$204.744$&$78.788$&$84.547$&$197.103$&$339.416$\\\hline
\end{tabular}
\end{table}

\begin{table}[htp!]
\caption{\small Hypothesis testing on the mixture weights of LGW model} \label{testpLGW}
\begin{tabular}{lcc} \hline
Hipótese&e-valor&p-valor$^*$\\\hline
$p_1=0$&$0.009$& $ 0.000$\\
$p_2=0$&$0.656$&$0.119$\\
$p_3=0$&$0.878$&$ 0.273$\\\hline
\multicolumn{3}{l}{\small *p-value calculated according to \citet{Diniz12}}
\end{tabular}
\end{table}

\begin{figure}[htp!]
\begin{center}
\includegraphics[scale=0.45]{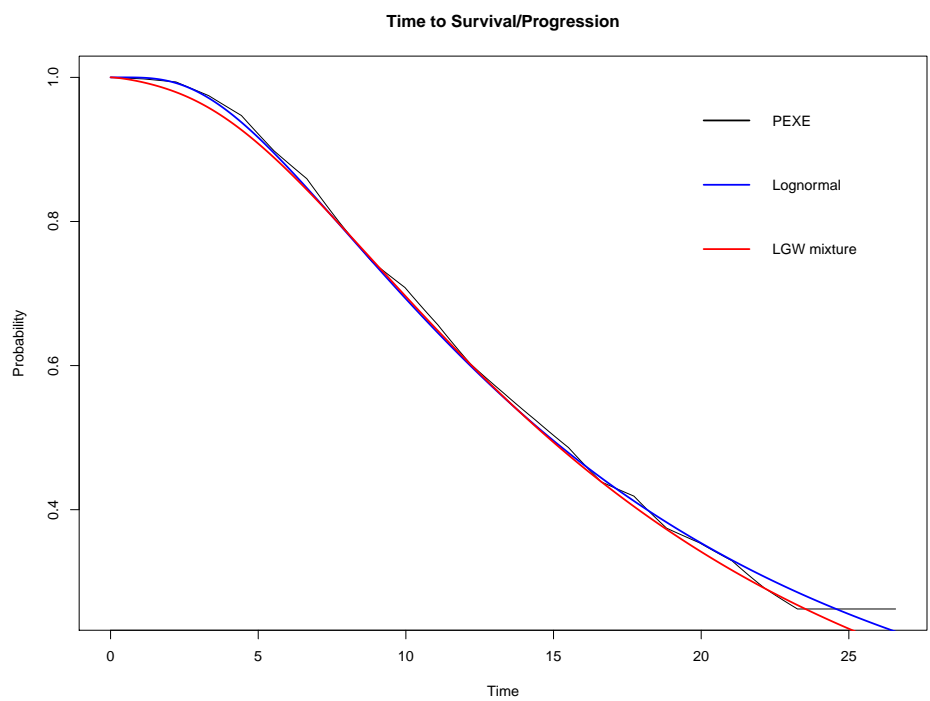}
\end{center}
\caption{Survival curves based on the estimates of the lognormal model, the LGW model and the PEXE}\label{LGWcurv}
\end{figure}

Figure \ref{LGWcurv} displays the survival curves calculated using Bayesian estimates of the lognormal model (Table \ref{LEst}), the LGW mixture model (Table \ref{EstLGW}) and a procedure called the piecewise exponential estimator (PEXE), introduced by \citet{Kim91}, representing the observed data. Unlike the well-known Kaplan-Meier estimator, the PEXE is smooth and continuous estimator of the survival function.

It appears reasonable to disregard both the gamma and the Weibull models; the lognormal model by itself produces a good estimate of survival function.

\begin{table}[htp!]
\caption{\small Summary of the posterior distribution of lognormal parameters}\label{LEst}
\begin{tabular}{lcccccc} \hline
Parâmetro&Mean&SD&$2.5\%$&Median&$97.5\%$ \\\hline
$\mu$&$20.298$ &$1.543$&$17.493$&$20.251$&$23.285$\\
$\sigma^2$&$355.367$&$101.927$&$191.546$&$343.222$&$589.779$\\\hline
\end{tabular}
\end{table}

Note that the preference for the lognormal model is evident in evaluating the LGW mixture model more than in the comparison between the lognormal and gamma distributions, where the evidence measures in favor of both models are very close. It means that the discrimination power provided by LGW model is much higher than the power of the pairwise comparisons. This finding is in agreement with the discussion of  \citet{Sawyer84}.

\vspace*{0.0 cm}
\section{Final Remarks}\label{sec7}
In this paper we considered the FBST for discriminating between survival distributions in the context of linear mixture model. The mixture approach allows us to compare between all alternative models at once by testing the hypotheses on the mixture weights space. The families of survival distributions considered include the lognormal, gamma and Weibull models. In this work, the density functions of the mixture components were reparametrized in terms of the mean $\mu$ and the variance $\sigma^2$ of the population so that all models under discrimination share common parameters \citep{kamary14,Pereira17}.

From the simulation results, we observed that the FBST achieves good performance on identifying the true distribution used to generate the survival times.

The application of the LGW mixture model to the survival data for HD patients allowed us to identify the lognormal distribution as the most appropriate in modeling observed data. Therefore, one can construct a regression model to the HD data considering the lognormal model as the distribution of the response variable.

It would be of interesting to apply the proposed procedure to survival data also considering another censoring mechanisms.
\section*{Acknowledgements}
The authors are grateful for the support of CNPq, COPPE/UFRJ and IME/USP.


\begin{thebibliography}{20}

\bibitem[\protect\citeauthoryear{Alves \textit{et al.}} {2014}]{Mauro}
Alves, M. and Souza e Silva, N. A. and Salis, L. H. A. and Pereira, B. B. and Godoy, P. H. and Nascimento, E. M. and Oliveira, J. M. F. (2014)
 Survival and Predictive Factors of Lethality in Hemodyalisis: D/I Polymorphism of The Angiotensin I-Converting Enzyme and of the Angiotensinogen M235T Genes. \textit{Arq Bras Cardiol.},
\textbf{103}, 209--218.

\bibitem[\protect\citeauthoryear{Araujo \textit{et al.}} {2005}]{Araujo05}
Araujo, M. I. and Pereira, B. B. and Cleroux, R. and Fernandes, M. and Lazraq, A. (2005)
Separate families of models: Sir David Cox contributions and recent developments. \textit{Student},
\textbf{5}, 251--258.

\bibitem[\protect\citeauthoryear{Araujo and Pereira}{2007}]{Araujo07}
Araujo, M. I. and Pereira, B. B. (2007)
A Comparison of Bayes Factors for Separated Models: Some Simulation Results. \textit{Communications in Statistics--Simulation and Computation},
\textbf{36}, 297--309.

\bibitem[\protect\citeauthoryear{Cox}{1961}]{Cox61}
Cox, D. R. (1961)
Tests of separate families of hypotheses. \textit{Proceedings 4th Berkeley Symposium in Mathematical Statistics and Probability},
\textbf{1}, 105--123.

\bibitem[\protect\citeauthoryear{Cox}{1962}]{Cox62}
Cox, D. R. (1962)
Further results on test of separate families of hypotheses. \textit{Journal of the Royal Statistical Society},
\textbf{B}, 406--424.

\bibitem[\protect\citeauthoryear{Cox}{1977}]{Cox77}
Cox, D. R. (1977)
The role of significance tests. \textit{Scand. J. Statist},
\textbf{4}, 49--70.

\bibitem[\protect\citeauthoryear{Diniz \textit{et al.}}{2012}]{Diniz12}
Diniz, M. and Pereira, C. A. B and Polpo, Adriano and Stern, J. M. and Wechsler, S. (2012)
Relationship between Bayesian and Frequentist significance indices. \textit{International Journal for Uncertainty Quantification},
\textbf{2}, 161--172.

\bibitem[\protect\citeauthoryear{Fletcher and Reeves}{1964}]{Fletcher}
Fletcher, R. and Reeves, C. M.  (1964)
Function minimization by conjugate gradients. \textit{Computer Journal},
\textbf{7}, 148--154.

\bibitem[\protect\citeauthoryear{Klein and Moeschberger}{2003}]{Klein}
Klein, J. and Moeschberger, M. L. (2003)
\textit{Survival analysis: techniques for censored and truncated data}, 2nd ed. Springer.

\bibitem[\protect\citeauthoryear{Haario \textit{et al.}}{2001}]{Haario01}
Haario, H. and Saksman, E. and Tamminen, J. (2001)
An adaptive Metropolis algorithm. \textit{Bernoulli},
\textbf{7}, 223--242.

\bibitem[\protect\citeauthoryear{Kamary \textit{et al.}}{2014}]{kamary14}
Kamary, K. and Mengersen, K. and Robert, C.P. and Rousseau, J. (2014) Testing hypotheses via a mixture estimation model. \textit{arXiv:1412.2044v2}.

\bibitem[\protect\citeauthoryear{Kempthorne }{1976}]{Kempt76}
Kempthorne, O. (1976)
Of what use are tests of significance and tests of hypothesis. \textit{Communications in Statistics -Theory and Methods},
\textbf{8}, 763--777.

\bibitem[\protect\citeauthoryear{Kim and Proschan}{1976}]{Kim91}
Kim, J. S. and  Proschan, F. (1991)
Piecewise exponential estimator of the survivor function. \textit{IEEE Transactions on Reliability},
\textbf{40}, 134--139.

\bibitem[\protect\citeauthoryear{Lauretto \textit{at al.}}{2003}]{Lauretto03}
Lauretto, M. and Pereira, C. A. B. and Stern, J. M. and Zacks, S. (2003) Comparing parameters of two bivariate normal distributions using the invariant full Bayesian significance test. \textit{Brazilian Journal of Probability and Statistics},
\textbf{17}, 147--168.

\bibitem[\protect\citeauthoryear{Lauretto and Stern}{2005}]{Lauretto05}
Lauretto, M. S. and Stern, J. M. (2005) FBST for mixture model selection.
 \textit{AIP Conference Proceedings},
\textbf{803}, 121--128.

\bibitem[\protect\citeauthoryear{Lauretto \textit{at al.}}{2007}]{Lauretto07}
Lauretto, M. S. and Faria Jr, S. R. and Pereira, C. A. B.and Pereira, B. B. and Stern, J. M. (2007) The problem of separate hypotheses via mixture models.
 \textit{AIP Conference Proceedings},
\textbf{954}, 268--275.

\bibitem[\protect\citeauthoryear{Lawless}{2002}]{Lawless02}
Lawless, J. F. (2002) \textit{Statistical Models and Methods for Lifetime Data}, 2nd ed. John Wiley \& Sons.

\bibitem[\protect\citeauthoryear{Lee and Wang}{2003}]{LeeWang}
Lee, E. T. and Wang, J. W. (2003) \textit{Statistical Methods for Survival Data analysis}, 3rd ed. Wiley.

\bibitem[\protect\citeauthoryear{Pereira}{1981}]{Pereira81}
Pereira, B. B. (1981) Choice of a survival model for patients with a brain tumour. \textit{Metrika},
\textbf{28}, 53--61.

\bibitem[\protect\citeauthoryear{Pereira and Pereira}{2017}]{Pereira17}
Pereira, B. B. and Pereira, C. A. B. (2017) \textit{Model Choice in Nonnested Families}, 1st edn. Berlin: Springer.

\bibitem[\protect\citeauthoryear{Pereira and Stern}{1999}]{PereiraS}
Pereira, C. A. B. and Stern, J. (1999) Evidence and Credibility: full Bayesian significance test for precise hypotheses. \textit{Entropy},
\textbf{1}, 69--80.

\bibitem[\protect\citeauthoryear{Pereira \textit{et al.}}{2008}]{PereiraSW}
Pereira, C. A. B. and Stern, J. and Wechsler, S. (2008) Can a significance test be genuinely Bayesian. \textit{Bayesian Analysis},
\textbf{3}, 79--100

\bibitem[\protect\citeauthoryear{Sawyer}{1984}]{Sawyer84}
Sawyer, K. R. (1984) Multiple hypotheses testing. \textit{Journal of teh Royal Statistical},
\textbf{Society-B 46}, 419--424.

\bibitem[\protect\citeauthoryear{Statisticat, LCC}{2016}]{LCC16}
Statisticat, LCC (2016) \textit{LaplacesDemon: A Complete Environment for Bayesian Inference within R}. R Package version 17.07.2016. https://cran.r-project.org/web/packages/LaplacesDemon/LaplacesDemon.pdf.

\bibitem[\protect\citeauthoryear{Stern and Pereira}{2014}]{SternP14}
Stern, J. and Pereira, C. A. B. (2014) Bayesian epistemic values: focus on surprise, measure probability. \textit{Logic Journal of The IGPL},
\textbf{22}, 236--254.

\bibitem[\protect\citeauthoryear{Wan}{2017}]{Wan17}
Wan, F. (2017) Simulating survival data with predefined censoring rates for proportional hazards models. \textit{Statistics in Medicine},
\textbf{36}, 838--854.

\end{thebibliography}

\end{document}